\documentclass{emulateapj}

\usepackage{amssymb}
\usepackage{graphicx}

\slugcomment{Accepted by ApJ, September 28th, 2007}
\shorttitle{Elemental Abundances of BCDs}
\shortauthors{Wu et al.}

\begin{document}

\title{Elemental Abundances of Blue Compact Dwarfs from mid-IR Spectroscopy with Spitzer}

\author{Yanling Wu\altaffilmark{1}, J. Bernard-Salas\altaffilmark{1},
  V. Charmandaris\altaffilmark{2,3}, V.  Lebouteiller\altaffilmark{1},
  Lei Hao\altaffilmark{1}, B.~R. Brandl\altaffilmark{4}, J.~R. Houck\altaffilmark{1}}

\altaffiltext{1}{Astronomy Department, Cornell University, Ithaca, NY 14853}

\altaffiltext{2}{University of Crete, Department of Physics, GR-71003,
Heraklion, Greece}

\altaffiltext{3}{IESL/Foundation for Research and Technology - Hellas,
  GR-71110, Heraklion, Greece and Chercheur Associ\'e, Observatoire de
  Paris, F-75014, Paris, France}

\altaffiltext{4}{Leiden Observatory, Leiden University, P.O. Box 9513,
2300 RA Leiden, The Netherlands}

\email{wyl@astro.cornell.edu, jbs@isc.astro.cornell.edu,
  vassilis@physics.uoc.gr, vianney@isc.astro.cornell.edu, 
  haol@astro.cornell.edu, brandl@strw.leidenuniv.nl,
  jrh13@cornell.edu}

\begin{abstract}
  We present a study of elemental abundances in a sample of thirteen
  Blue Compact Dwarf (BCD) galaxies, using the $\sim$10--37$\mu$m high
  resolution spectra obtained with Spitzer/IRS. We derive the
  abundances of neon and sulfur for our sample using the infrared
  fine-structure lines probing regions which may be obscured by dust
  in the optical and compare our results with similar infrared studies
  of starburst galaxies from ISO.  We find a good correlation between
  the neon and sulfur abundances, though sulfur is under-abundant
  relative to neon with respect to the solar value. A comparison of
  the elemental abundances (neon, sulfur) measured from the infrared
  data with those derived from the optical (neon, sulfur, oxygen)
  studies reveals a good overall agreement for sulfur, while the
  infrared derived neon abundances are slightly higher than the
  optical values.  This indicates that either the metallicities of
  dust enshrouded regions in BCDs are similar to the optically
  accessible regions, or that if they are different they do not
  contribute substantially to the total infrared emission of the host
  galaxy.

\end{abstract}

\keywords{dust, extinction ---
  galaxies:starburst ---
  abundances}

\section{Introduction}

Blue Compact Dwarf Galaxies (BCDs) are dwarf galaxies with blue
optical colors resulting from one or more intense bursts of
star-formation, low luminosities (M$_B>-$18) and small sizes. The
first BCD discovered was I\,Zw\,18 by \citet{Zwicky66}, which had the
lowest oxygen abundance observed in a galaxy \citep{Searle72}, until
the recent study of the western component of SBS0335-052
\citep{Izotov05}.  Although BCDs are defined mostly by their
morphological parameters, they are globally found to have low heavy
element abundances as measured from their HII regions
(1/30\,Z$_\odot$$\sim$1/2\,Z$_\odot$). The low metallicity of BCDs is
suggestive of a young age since their interstellar medium is
chemically unevolved. However, some BCDs do display an older stellar
population and have formed a large fraction of their stars more than
1Gyr ago \citep[see][]{Loose85,Aloisi07}. The plausible scenario that
BCDs are young is intriguing within the context of Cold Dark Matter
models which predict that low-mass dwarf galaxies, originating from
density perturbations much less massive than those producing the
larger structures, can still be forming at the current epoch. However,
despite the great success in detecting galaxies at high redshift over
the past few years, bona fide young galaxies still remain extremely
difficult to find in the local universe
\citep{Kunth86,Kunth00,Madden06}. This is likely due to the
observational bias of sampling mostly luminous more evolved galaxies
at high redshifts. If some BCDs are truly young galaxies, they would
provide an ideal local laboratory to understand the galaxy formation
processes in the early universe.

Over the past two decades, BCDs have been studied extensively in many
wavelengths using ground-based and space-born instruments \citep [for
a review see][]{Kunth00}.  In the FUV, the Far Ultraviolet
Spectroscopic Explorer (FUSE) has been used to study the chemical
abundances in the neutral gas in several BCDs \citep{Thuan02,
  Aloisi03, Lebouteiller04}. Optical spectra have been obtained for a
large number of BCDs and display strong narrow emission lines
resulting from the intensive star-formation processes that take place
in these systems \citep{Izotov97, Izotov99b, Pustilnik05, Salzer05}.
The Infrared Space Observatory (ISO) revealed unexpectedly that
despite their low metallicities, BCDs, such as SBS\,0335-052E, could
still have copious emission from dust grains \citep{Thuan99, Madden00,
  Madden06, Plante02}. More recently, the Spitzer Space Telescope
\citep{Werner04} has been used to observe these metal-poor dwarf
systems in order to study their dust continuum properties and the
polycyclic aromatic hydrocarbon (PAH) features \citep{Houck04b,
  Hogg05, Engelbracht05, Rosenberg06, Wu06, OHalloran06, Hunt06,
  Wu07}. Finally, radio observations have also been performed for
several BCDs to study their HI kinematics and distribution
\citep{Thuan04} and thermal/non-thermal continuum emission properties
\citep{Hunt05}.

\begin{deluxetable*}{lrllllcc}
  \tabletypesize{\scriptsize}
  \setlength{\tabcolsep}{0.02in}
  \tablecaption{Observing Parameters of the Sample\label{tab1}}
  \tablewidth{0pc}
  \tablehead{
    \colhead{Object Name} & \colhead{RA} & \colhead{Dec} & \colhead{AORKEY} &
    \colhead{Observation} & \colhead{Redshift} & \multicolumn{2}{c}{On-source Time (Sen)}\\
    \colhead{} & \colhead{(J2000)} & \colhead{(J2000)} & \colhead{} & \colhead{Date} &  \colhead{} & 
    \colhead{SH} & \colhead{LR} \\
  }
  \startdata
  
Haro11              & 00h36m52.5s  & -33d33m19s &  9007104 & 2004-07-17  &  0.0206 & 480     & 240      \\
NGC1140             & 02h54m33.6s  & -10d01m40s &  4830976 & 2004-01-07  &  0.0050 & 480     & 240      \\
SBS0335-052E        & 03h37m44.0s  & -05d02m40s & 11769856 & 2004-09-01  &  0.0135 & 1440    & 960      \\
NGC1569             & 04h30m47.0s  & +64d50m59s &  9001984 & 2004-03-01  & $\sim$ 0& 480     & 240      \\
IIZw40              & 05h55m42.6s  & +03d23m32s &  9007616 & 2004-03-01  &  0.0026 & 480     & 240      \\
UGC4274             & 08h13m13.0s  & +45d59m39s & 12076032 & 2004-10-23  &  0.0015 & 120     &  56      \\ 
                    &              &            & 12626688 & 2004-11-11  &         & 120     &  56      \\ 
IZw18               & 09h34m02.0s  & +55d14m28s &  9008640 & 2004-03-27  &  0.0025 & 480     & 240      \\
                    &              &            & 16205568 & 2005-12-16  &         & 2880    & 1440     \\
VIIZw403            & 11h27m59.9s  & +78d59m39s &  9005824 & 2004-12-09  & $\sim$ 0& 480     & 240      \\
Mrk1450             & 11h38m35.6s  & +57d52m27s &  9011712 & 2004-12-12  &  0.0032 & 480     & 240      \\
UM461               & 11h51m33.3s  & -02d22m22s &  9006336 & 2005-01-03  &  0.0035 & 480     & 240      \\
                    &              &            & 16204032 & 2006-01-14  &         & 1440    & \nodata  \\
SBS1210+537A        & 12h12m55.9s  & +53d27m38s &  8989952 & 2004-06-06  & \nodata & 480     & 240      \\
Tol1214-277         & 12h17m17.1s  & -28d02m33s &  9008128 & 2004-06-28  &  0.0260 & 480     & 240      \\
Tol65               & 12h25m46.9s  & -36d14m01s &  4829696 & 2004-01-07  &  0.0090 & 480     & 240      \\
UGCA292             & 12h38m40.0s  & +32d46m01s &  4831232 & 2004-01-07  &  0.0010 & 480     & 240      \\
Tol1304-353         & 13h07m37.5s  & -35d38m19s &  9006848 & 2004-06-25  &  0.0140 & 480     & 240      \\
Pox186              & 13h25m48.6s  & -11d37m38s &  9007360 & 2004-07-14  &  0.0039 & 480     & 240      \\
CG0563              & 14h52m05.7s  & +38d10m59s &  8992512 & 2005-05-30  &  0.0324 & 240     & 120      \\
CG0598              & 14h59m20.6s  & +42d16m10s &  8992256 & 2005-03-19  &  0.0575 & 480     & 240      \\
CG0752              & 15h31m21.3s  & +47d01m24s &  8991744 & 2005-03-19  &  0.0211 & 480     & 240      \\
Mrk1499             & 16h35m21.1s  & +52d12m53s &  9011456 & 2004-06-05  &  0.0090 & 480     & 240      \\
{\rm [RC2]}A2228-00 & 22h30m33.9s  & -00d07m35s &  9006080 & 2004-06-24  &  0.0052 & 480     & 240      \\

\enddata

\tablecomments{The coordinates and redshifts of the objects are cited from the NASA/IPAC Extragalactic Database (NED).
  In this paper, we only include the analysis of thirteen out of twenty-two sources which have SNRs sufficient for our 
  abundance study. CG0563, CG0598 and CG0752 are included in the original sample as BCD candidates, however, they appear 
  to be more starburst like \citep[see][]{Hao07}. Thus even though they have high SNR, we do not include these three sources 
  in this study.}
\end{deluxetable*}

Metallicity is a key parameter that influences the formation and
evolution of both stars and galaxies. Detailed studies of the
elemental abundances of BCDs have already been carried out by several
groups \citep{Izotov99b, Kniazev03, Shi05} and the well known
metallicity-luminosity relation has also been studied in detail in the
environment of dwarf galaxies \citep{Skillman89, Hunter99,
  Melbourne02}. However, because these studies were performed in the
optical, they were limited by the fact that the properties of some of
the deeply obscured regions in the star-forming galaxies may remain
inaccessible due to dust extinction at these wavelengths. In fact,
\citet{Thuan99}, using ISO, have shown that the eastern component of
SBS\,0335-052 does have an embedded super star cluster (SSC) that is
invisible in the optical while contributing $\sim$75\% to the
bolometric luminosity \citep[see also][]{Plante02,Houck04b}, even
though it has very low metallicity (12+log(O/H)=7.33), which would in
principle imply a low dust content. In addition to probing the dust
enshrouded regions, emission in the infrared also has the advantage
that the lines accessible at these wavelengths are less sensitive to
the electron temperature fluctuations than the corresponding optical
lines of the same ion. In the infrared, more ionization stages of an
element become available as well. The improved sensitivity of the
Infrared Spectrograph (IRS\footnote{The IRS was a collaborative
  venture between Cornell University and Ball Aerospace Corporation
  funded by NASA through the Jet Propulsion Laboratory and the Ames
  Research Center.})  \citep{Houck04a} on Spitzer has enabled us to
obtain for the first time infrared spectra for a much larger sample of
BCDs than was previously possible \citep{Thuan99, Madden00, Verma03,
  Martin06}, thus motivating this study to probe the heavy element
abundances in BCDs.

In this paper, we analyze Spitze/IRS spectra of thirteen BCDs and
present elemental abundances of neon and sulfur, which are both
primary elements produced by the same massive stars in the nuclear
synthesis processes. In section 2, we describe the sample selection,
observations and data reduction. We present our results on the
chemical abundances in section 3, followed by a comparison of the
optical and infrared derived abundances in section 4. We show the
interplay between the abundances and PAH emission in section 5.
Finally, we summarize our conclusions in section 6.

\section{Observations and Data Reduction}

As part of the {\em IRS} Guaranteed Time Observation (GTO) program, we
have compiled a large sample of BCD candidates selected from the
Second Byurakan Survey (SBS), Bootes void galaxies
\citep{Kirshner81,Popescu00}, and other commonly studied BCDs. Details
on the low-resolution spectra of the sample have been published by Wu
et al. (2006).

We acquired the targets using the red (22$\mu$m) IRS peak-up camera in
high accuracy mode to locate the mid-IR centroid of the source and
then offset to the appropriate slit using the standard IRS staring
observing mode. Subsequently, we obtained the 10$-$37\,$\mu$m spectra
for our sources using the {\em IRS} Short-High (SH, 9.9-19.6\,$\mu$m)
and Long-high (LH, 18.7-37.2\,$\mu$m) modules. No spectra of the
background were obtained. The AORkey and on-source integration time of
the twenty-two sources for which we obtain high-resolution
spectroscopy are given in Table \ref{tab1}.  In this paper, we focus
on the description and analysis of thirteen BCDs from our program
which have a signal-to-noise ratio (SNR) high enough for our elemental
abundance study.

The data were processed by the {\em Spitzer} Science Center (SSC)
pipeline version 13.2. The three-dimensional data cubes were converted
to two-dimensional slope images after linearization correction,
subtraction of darks, and cosmic-ray removal. The reduction of the
spectral data started from intermediate pipeline products ``droop''
files, which only lacked stray light and flat-field correction.
Individual pointings to each nod position of the slit were co-added.
The data from SH and LH were extracted using the full slit extraction
method of a script version of {\em IRS} data analysis package SMART
\citep{Higdon04} from the median of the combined images. The
1-dimensional spectra were flux calibrated by multiplying by a
relative spectral response function (RSRF), which was created from the
{\em IRS} standard star, $\xi$ Dra, for which accurate templates were
available \citep{Cohen03,Sloan07}. As a final cosmetic step the ends
of each order where the noise increases significantly were manually
clipped. No scaling was needed between the adjacent orders within the
same module. For faint sources, the spectra taken at nod position 1
are severely affected by fringing problems, thus we only use the data
at nod position 2 for these sources in our study.

The fine-structure lines, [SIV]\,$\lambda$10.51$\mu$m,
[NeII]\,$\lambda$12.81$\mu$m, [NeIII]\,$\lambda$15.55$\mu$m and
[SIII]\,$\lambda\lambda$18.71$\mu$m, 33.48$\mu$m are clearly present
in the majority of the BCDs in our sample\footnote {See Fig. 3 of
  \citet{Wu06}, or \citet{Hao07} for the reduced spectra.}. We measure
the line fluxes by fitting them with a Gaussian profile. Even though
the [SIII] line is visible in both SH and LH we only use the
18.71\,$\mu$m line for deriving the ionic abundance. This is because
by using lines within the same module (SH), we remove uncertainties
due to the different sizes of the SH and LH slits. Furthermore, the
[SIII]\,$\lambda$33.48$\mu$m line is near the cut off of the LH module
where the sensitivity drops dramatically, making its measurement more
uncertain in some cases.

\section{Elemental Abundances of BCDs}                                        

In this section, we derive the neon and sulfur abundances using the
new infrared data of our BCD sample. The elemental abundances of the
compact HII regions in our Galaxy \citep{Martin02,Simpson04}, in the
Magellanic Clouds \citep{Vermeij02}, as well as in starburst galaxies
\citep{Verma03, Martin06} have already been studied extensively in the
past using ISO, and more recently with {\em Spitzer}
\citep{Rubin07}.  There are several methods to derive the chemical
abundances \citep [for a review see][]{Stasinska07}. Here we use an
empirical method, which derives ionic abundances directly from the
observed lines of the relavent ions.  To do so, we need to have the
flux of at least one hydrogen recombination line, the dust extinction,
as well as the electron temperature and density of the interstellar
medium (ISM).

\subsection{Electron density and temperature}

The electron density (N$_e$) could in principle be determined by
comparing the measured ratio of [SIII]18.71$\mu$m/33.48$\mu$m to the
expected theoretical value using the corresponding $S$-curve
\citep[see][]{Houck84}. However, the values for the ratio are in the
horizontal part of the $S$-curve and as a result we cannot use these
two lines to accurately constrain the electron density of these
systems.  Furthermore, the 33.48\,$\mu$m [SIII] line is located at the edge
of the LH slit, where the sensitivity drops dramatically, and
considering that these emission lines are weak, the measured line flux
for the 33.48\,$\mu$m [SIII] has a large uncertainty. Since the infrared
determination of elemental abundances does not depend strongly on the
density, we adopted the optically derived electron densities from the
literature, which range from 10 to 3000\,cm$^{-3}$ (see Table \ref
{tab2}). For sources where such information is not available, we adopt
a typical electron density of 100\,cm$^{-3}$ in this paper.

\begin{deluxetable*}{lccccccrc}
  \tabletypesize{\scriptsize}
  \setlength{\tabcolsep}{0.02in}
  \tablecaption{Optical Properties of the Sources\label{tab2}} 
  \tablewidth{0pc}
  \tablehead{
    \colhead{Object} & \multicolumn{4}{c}{F(H$\beta$) ($\times$10$^{-14}$ergs~s$^{-1}$cm$^{-2}$)} &
    \colhead{E$_{B-V}$} & \colhead{T$_e$} & \colhead{N$_e$} & \colhead{ref} \\
    \cline{2-5}
    \colhead{} &  \colhead{Hu$\alpha$-derived} & \colhead{H$\alpha$-derived} & \colhead{radio-derived} &
    \colhead{optical} & \colhead{mag} & \colhead{K} & \colhead{cm$^{-3}$} & \colhead{} \\
  }
  \startdata
  
  Haro11        & 252$\pm$40  & \nodata    & \nodata    & \nodata    & 0.41  & 13700 & 10  & (1)\\
  NGC1140       &  67$\pm$9   & \nodata    & \nodata    & \nodata    & 0.10  & 10000 & 100 & (2)\tablenotemark{a} \\
  SBS0335-052E  & $<$22.4     & \nodata    &6.5$\pm$0.06& \nodata    & 2.76  & 20000 & 200 & (3),(4),(5)\\
  NGC1569       & 393$\pm$19  & \nodata    & \nodata    & \nodata    & 0.65  & 12000 & 100 & (6)\\ 
  IIZw40        & 363$\pm$10  & 282$\pm$22 & \nodata    & \nodata    & 0.79  & 13000 & 190 & (2),(7) \\
  UGC4274       & $<$59.6     &11.5$\pm$1.0& \nodata    & \nodata    & 0     & 10000 & 62  & (7),(8)\tablenotemark{a}\\
  IZw18         & $<$12.6     & 8.0$\pm$0.6& 6.1$\pm$0.6& \nodata    & 0.08  & 19000 & 100 & (7),(9),(10)\\
  VIIZw403      & $<$24.6     &11.8$\pm$1.2& \nodata    & \nodata    & 0     & 14800 & 100 & (7),(11)\\
  Mrk1450       & $<$14.8     &15.8$\pm$1.4& \nodata    & \nodata    & 0.10  & 12500 & 100 & (7),(12)\\
  UM461         & $<$21.0     & \nodata    & \nodata    &13.6$\pm$0.6& 0.08  & 16100 & 200 & (13),(14)\\
  Tol1214-277   & 23.3$\pm$2.8& \nodata    & \nodata    & \nodata    & 0.03  & 19790 & 400 & (15)\\  
  Tol65         & $<$18.2     & 9.3$\pm$0.4& \nodata    & \nodata    & 0.08  & 17320 & 50  & (7),(15)\\ 
  Mrk1499       & 14.6$\pm$1.9&15.1$\pm$1.3& \nodata    & \nodata    & 0.17  & 12600 & 3267& (7),(16)\\ 
  \enddata

  \tablecomments{When no data are available in the literature, the 
    electron temperature and densities are assumed to be 10\,000\,K (
    for NGC1140 and UGC4274) and 100~cm$^{-3}$ (for NGC1140 and NGC1569) 
    respectively. When the Hu\,$\alpha$ line is detected, the H$\beta$
    flux is preferentially derived from this line. The upper limits 
    are listed for the non-detections. For the remaining sources,
    we derived the H$\beta$ flux either from the thermal component of the
    radio continuum, or from the extinction corrected H$\alpha$ flux
    inside the SH slit. For UM461, the H$\beta$ flux is for the whole 
    galaxy from the integrated optical spectra.}  
  \tablerefs{(1) \citet{Bergvall02}, (2) \citet{Guseva00},
    (3) \citet{Izotov06}, (4) \citet{Hunt04}, (5) \citet{Houck04b},
    (6) \citet{Kobulnicky97}, (7) \citet{Gildepaz03}, (8)
    \citet{Ho97}, (9) \citet{Cannon05}, (10) \citet{Izotov99a}, (11)
    \citet{Izotov97}, (12) \citet{Izotov94}, (13) \citet{Moustakas06},
    (14) \citet{Izotov98}, (15) \citet{Izotov01}, (16) \citet{Kong02}
    }
   
\end{deluxetable*}

Since BCDs in general have low metallicities, it is expected that BCDs
would have relatively high electron temperatures (T$_e$). Fortunately,
the infrared lines are much less sensitive to the uncertainties in the
electron temperature compared to the optical \citep{Bernard-Salas01}.
As a result we adopt electron temperatures derived from the optical
studies found in the literature. For sources which do not have direct
measurements (which happen to be high metallicity sources in our
sample, e.g. NGC1140, UGC4274), we use a representative temperature of
T$_e$\,=\,10,000\,K (see Table \ref {tab2}).

\subsection{Ionized hydrogen flux estimates}

In order to derive the ionic abundances, we also need to obtain
information on at least one hydrogen recombination line (usually
H$\beta$). For sources in which the Humphreys\,$\alpha$ (Hu$\alpha$,
12.37\,$\mu$m) line is detected, we use it to convert to H$\beta$.
From the extinction corrected Hu$\alpha$ flux, we estimate the
H$\beta$ flux using the values given in the tables of \citet{Hummer87}
for Case B recombination. This has the advantage that we do not need
to correct for aperture effects because the neon and sulfur lines we use
to derive the abundances also reside in the same module (SH). The
12.37\,$\mu$m line is generally very weak and it is detected in six of
our thirteen sources. For SBS0335-052E and IZw18, which have radio
continuum data, we calculate the H$\beta$ flux from the thermal
free-free emission using the fraction included in the aperture of our
SH slit at the time of the observation, and no correction for
extinction is needed. For the remaining four sources which have
H$\alpha$ images available from \citet{Gildepaz03}, we overlay the SH
slit on each H$\alpha$ image to calculate what fraction of the
H$\alpha$ emission is included in our slit. Then we derive the
corresponding H$\beta$ flux inside the SH slit from the
extinction-corrected H$\alpha$ flux. Finally, for UM461, we use the
H$\beta$ flux of the whole galaxy from the integrated spectra
\citep{Moustakas06}. UM461 is a very compact source, thus fully
encompassed within the SH slit.

\subsection{Extinction correction}

All our measurements, both the fluxes from the H$\alpha$ images, as
well as the infrared lines, have been corrected for extinction. From
the low-resolution spectra of our BCDs \citep{Wu06}, we find that most
of the sources do not show a strong 9.7\,$\mu$m silicate feature, thus
indicating an intrinsically relatively low dust extinction. Due to the
faintness of our targets at $\lambda<$9.7\,$\mu$m, the determination
of the continuum on the blue side of the 9.7\,$\mu$m feature is often
poor and the uncertainties in estimating the mid-IR extinction from
this are rather large \citep[see discussion in][]{Spoon07}. Thus for
sources where optical spectra are available, we adopt the E$_{\rm
  B-V}$ values calculated from the hydrogen recombination lines in the
literature. Only for SBS\,0335-052E, which has a high quality mid-IR
spectrum and evidence for an embedded SSC, the optical E$_{B-V}$ is
not a good estimate of its extinction, thus we use the extinction
estimated from the depth of the silicate feature \citep{Houck04b}.
Throughout this paper, we adopt the \citet{Fluks94} extinction law,
though using the \citet{Draine03} law would produce very similar
results. The E$_{\rm B-V}$ magnitudes are provided in Table
\ref{tab2}.

\subsection{Neon and sulfur  abundances determination}

We use the electron density and temperature given in Table \ref
{tab2}, together with the fine structure lines detected in the {\em
  IRS} high-resolution spectra, to calculate the abundances of neon
and sulfur. The equation used to determine the ionic abundance is
described in \citet{Bernard-Salas01} and it is:

\begin{equation}
  \frac{N_{\mathrm{ion}}}{N_{\mathrm{p}}}= \frac{I_{\mathrm{ion}}}{I_
    {\mathrm{H_{\beta}}}} N_{\mathrm{e}}
  \frac{\lambda_{\mathrm{ul}}}{\lambda_{\mathrm{H_{\beta}}}} \frac
  {\alpha_{\mathrm{H_{\beta}}}}{A_{\mathrm{ul}}}
  \left( \frac{N_{\mathrm{u}}}{N_{\mathrm{ion}}} \right)^{-1}
  \label{eq_abun_ch2}
\end{equation}

where I$_{\mathrm{ion}}$/I$_{\mathrm{H\beta}}$ is the measured flux of
the fine-structure line normalized to H$\beta$; N$_{\mathrm{P}}$ is
the density of the ionized hydrogen; $\lambda_{\mathrm{ul}}$ and
$\lambda_{\mathrm{H\beta}}$ are the wavelengths of the line and
H$\beta$; $\alpha_{\mathrm{H\beta}}$ is the effective recombination
coefficient for H$\beta$; A$_{\mathrm{ul}}$ is the Einstein
spontaneous transition rate for the line and N$_\mathrm
u$/N$_{\mathrm{ion}}$ is the ratio of the population of the level from
which the line originates to the total population of the ion. This
ratio is determined by solving the statistical equilibrium equation
for a five level system and normalizing the total number of ions to be
unity \citep{Osterbrock89}. The effective collisional strengths used
to derive the population of levels were obtained from the appropriate
reference of the IRON project
\citep{Hummer93}\footnote{http://www.usm.uni-muenchen.de/people/ip/iron-project.html}.

\begin{deluxetable*}{lrrrrrrrrr}
\tabletypesize{\scriptsize}
  \setlength{\tabcolsep}{0.02in}
  \tablecaption{Fine-structure Line Fluxes and Ionic Abundances\label{tab3}}
  \tablewidth{0pc}
  \tablehead{
    \colhead{} & \multicolumn{4}{c}{Line Fluxes ( $\times10^{-14}$ergs~cm$^{-2}$~s$^{-1}$)} & \colhead{} &\multicolumn{4}{c}{Ionic Abundances ($\times10^{-6}$)}\\
    \cline{2-5}\cline{7-10}
   \colhead{Object} & \colhead{[SIV](10.51$\mu$m)} & \colhead{[NeII](12.81$\mu$m)} & \colhead{[NeIII](15.55$\mu$m)} 
    & \colhead{[SIII](18.71$\mu$m)} & \colhead{} & \colhead{N$_{\rm SIV}$/N$_p$} & \colhead{N$_{\rm NeII}$/N$_p$} & \colhead{N$_{\rm NeIII}$/N$_p$}& 
    \colhead{N$_{\rm SIII}$/N$_p$} \\
  }
  \startdata  
  Haro11        & 43.95$\pm$0.14  &   31.23$\pm$0.23  & 98.14$\pm$0.65  &   45.24$\pm$0.25  & & 0.353$\pm$0.056 &    15.4$\pm$2.4   &  23.2$\pm$3.7   &    1.89$\pm$0.30  \\
  NGC1140       & 12.14$\pm$0.13  &   11.41$\pm$0.20  & 37.79$\pm$0.32  &   20.04$\pm$0.11  & & 0.375$\pm$0.051 &    22.5$\pm$3.0   &  35.5$\pm$4.8   &    3.56$\pm$0.48 \\
  SBS0335-052E  &  1.56$\pm$0.03  &   $<$0.14         &  1.43$\pm$0.02  &    0.43$\pm$0.03  & & 0.527$\pm$0.011 &   $<$2.27         &  11.2$\pm$1.6   &   0.535$\pm$0.038 \\
  NGC1569       &147.26$\pm$0.55  &   15.70$\pm$0.17  &175.85$\pm$0.59  &   72.70$\pm$0.23  & & 0.817$\pm$0.040 &    5.28$\pm$0.26  &  28.2$\pm$1.4   &    2.12$\pm$0.10  \\
  IIZw40        &185.72$\pm$1.73  &    6.24$\pm$0.11  &112.65$\pm$0.83  &   45.80$\pm$0.21  & &  1.12$\pm$0.03  &    2.23$\pm$0.07  &  19.3$\pm$0.6   &    1.37$\pm$0.04  \\
  UGC4274       &  4.18$\pm$0.12  &    8.87$\pm$0.11  & 12.84$\pm$0.12  &   12.63$\pm$0.11  & & 0.743$\pm$0.068 &     102$\pm$9     &  70.3$\pm$6.1   &    13.2$\pm$1.2 \\  
  IZw18         &  0.48$\pm$0.03  &    0.09$\pm$0.01  &  0.46$\pm$0.02  &    0.23$\pm$0.02  & & 0.125$\pm$0.015 &    1.41$\pm$0.21  &  3.56$\pm$0.38  &   0.276$\pm$0.036 \\
  VIIZw403      &  0.99$\pm$0.05  &   $<$0.142        &  0.88$\pm$0.02  &    0.87$\pm$0.02  & & 0.311$\pm$0.035 &   $<$2.79         &  8.41$\pm$0.88  &    1.39$\pm$0.14  \\
  Mrk1450       &  7.34$\pm$0.04  &    1.34$\pm$0.03  &  9.61$\pm$0.04  &    4.37$\pm$0.03  & & 0.928$\pm$0.223 &   10.70$\pm$2.57  &  36.9$\pm$8.9   &    2.98$\pm$0.71  \\
  UM461         &  4.58$\pm$0.04  &    0.16$\pm$0.02  &  2.83$\pm$0.03  &   0.812$\pm$0.03  & & 0.636$\pm$0.029 &    1.38$\pm$0.18  &  11.8$\pm$0.5   &   0.563$\pm$0.032  \\
  Tol1214-277   &  0.88$\pm$0.02  &   $<$0.10         &  0.61$\pm$0.02  &   $<$0.19         & & 0.061$\pm$0.007 &   $<$0.413        &  1.23$\pm$0.15  &   $<$0.059         \\
  Tol65         &  0.69$\pm$0.02  &   $<$0.14         &  0.90$\pm$0.02  &    0.32$\pm$0.02  & & 0.129$\pm$0.007 &   $<$1.62         &  5.07$\pm$0.25  &   0.293$\pm$0.022 \\
  Mrk1499       &  1.74$\pm$0.02  &    1.88$\pm$0.03  &  5.03$\pm$0.03  &    2.91$\pm$0.02  & & 0.286$\pm$0.037 &    16.5$\pm$2.2   &  21.6$\pm$2.8   &    2.28$\pm$0.30  \\
  \enddata
  \tablecomments{The observed line fluxes are measured from the IRS high resolution spectra of the sources. Background emission has not been subtracted, but 
    this does not affect the flux of the fine-structure lines.}
\end{deluxetable*}

The most important ionization stages of neon and sulfur are available
in the infrared. We detected [SIV]\,$\lambda$10.51\,$\mu$m and
[NeIII]\,$\lambda$15.56\,$\mu$m in all of our BCDs while the
[NeII]\,$\lambda$12.81\,$\mu$m line is detected in nine objects and
[SIII]\,$\lambda$18.71\,$\mu$m in twelve sources (see Table \ref
{tab3}). For sulfur, SII has its strongest emission lines in the
optical.  \citet{Vermeij02} have shown that SII is typically less than
10\% of SIII in the Magellanic cloud HII regions they studied. Because
BCDs are typically high-excitation objects, we do not expect to have a
significant contribution from SII. The optical studies reveal that the
ionic abundance of SII is typically 10$-$20\% of that of SIII
\citep{Izotov94,Izotov97}, thus we keep in mind that this might result
in a $\sim$10\% underestimate in our infrared derived sulfur
abundance.  This is addressed further in the following subsection where
we discuss the properties of the individual objects. The presence of
the [OIV]\,$\lambda$25.89\,$\mu$m line in most of our sources (except for
VIIZw403, UM461, Tol65 and Mrk1499) indicates that some [NeIV] might
be present, even though the [OIV] line is typically weak. For example,
in Tol1214-277, where [OIV] is detected, the flux of
[NeIV]$\lambda$4725 from optical spectra is less than 1\% of that of
[NeIII]$\lambda$3868 \citep{Izotov04}, indicating that the
contribution of [NeIV] is not significant. \citet{Tsamis05} also have
shown in their photoionization modelling study of 30 Doradus that the
contribution from NeIV to the total neon abundance is much less than
1\%.  As a result, rather than applying ionization correction factors
(ICFs) which could introduce an unknown systematic uncertainty in our
study, we sum the corresponding ionic abundances for the most
important ionization stages to determine the total element abundance
and the results are shown in Table \ref{tab3}.

Two successive stages of ionization of a given element (X) can be used
to measure the state of the ionization of the ISM, which, on first
order, depends on the ionization parameter, U, and the hardness of the
ionizing radiation \citep{Vilchez88}. For a given U, the ratio
X$^{+i+1}$/X$^{+i}$ is an indicator of the hardness of the stellar
radiation field. Here we use the ionic pairs of Ne$^{++}$/Ne$^+$ and
S$^{+3}$/S$^{++}$ as indicators of the hardness of the ionization field
and plot them in Fig. 1. It is clear from the figure that these two
ratios increase proportionally.  This indicates that the hardening of
the radiation field affects similarly the various elements in the full
range of the ionizing continua.

\begin{figure}
  \epsscale{1.2}
  \plotone{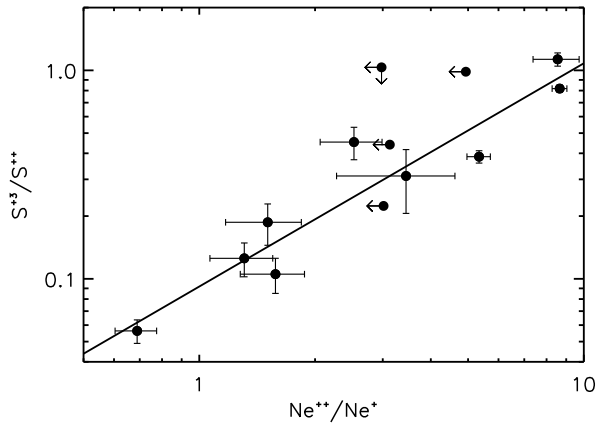}
  \caption{A plot of Ne$^{++}$/Ne$^+$ vs S$^{3+}$/S$^{++}$, which
    traces the hardness of the radiation field. The solid line is a
    fit to the data.}
  \label{fig:fig1}
\end{figure}

The major uncertainty in our study of the elemental abundances
originates from the H$\beta$ fluxes used as references. For sources
where the 12.37\,$\mu$m Hu$\alpha$ line is detected, we have no
uncertainties from matching the slit apertures, and the extinction
correction is very small. For sources where we estimate the H$\beta$
flux from the thermal radio continuum emission, we are free of
extinction corrections, even though we still need to properly account
for the radio emission which falls within the IRS slit. For BCDs with
no Hu\,$\alpha$ emission or radio data available in the literature, we
estimate the H$\beta$ flux using archival H$\alpha$ images of the objects.
This is clearly more challenging because in addition to the aperture
corrections, the extinction correction is higher which could introduce
a systematic uncertainty as large as 25\%. In the following analysis, we
indicate the sources with or without the detection of Hu\,$\alpha$
line using different symbols on the plots (see Fig. 2, 3 and 4). The
measurement uncertainties in the infrared lines are typically only a few
percent, but could grow to $\sim$10\% in the case of fainter sources.
The uncertainties we list in Table \ref{tab3} only account for measurement
uncertainties.

\begin{deluxetable*}{lrrrrrrr}
\tabletypesize{\scriptsize}
  \setlength{\tabcolsep}{0.02in}
  \tablecaption{Elemental Abundances\label{tab4}}
  \tablewidth{0pc}
  \tablehead{
    \colhead{Object} & \colhead{Ne/H$ (\times$10$^{-6}$)} &
  \colhead{S/H ($\times$10$^{-6}$)} & \colhead{O/H ($\times$10$^{-5}$)} & 
    \colhead{Ne (Z$_\odot$)} & \colhead{S (Z$_\odot$)} & \colhead{O (Z$_\odot$)} & \colhead{ref (O/H)} \\ 
  }
  \startdata  
  Haro11        &  38.6$\pm$4.4  & 2.24$\pm$0.31  & 7.94  &  0.33$\pm$0.04 & 0.16$\pm$0.02  & 0.16  & (1) \\
  NGC1140       &  58.0$\pm$5.7  & 3.94$\pm$0.48  & 28.8  &  0.48$\pm$0.05 & 0.28$\pm$0.03  & 0.59  & (2) \\
  SBS0335-052E  &  [11.2, 13.5]\tablenotemark{a}  & 1.06$\pm$0.04 & 1.95$\pm$0.09 &  [0.09, 0.11]\tablenotemark{a} & 0.08$\pm$0.003 & 0.03$\pm$0.001 & (3) \\
  NGC1569       &  33.5$\pm$1.4  & 2.94$\pm0.11$  & 15.8$\pm$0.5  &  0.28$\pm$0.01 & 0.21$\pm$0.01  & 0.32$\pm$0.01 & (4) \\
  IIZw40        &  21.5$\pm$0.6  & 2.49$\pm$0.05  & 12.2$\pm$0.4  &  0.17$\pm$0.005& 0.18$\pm$0.004 & 0.25$\pm$0.01 & (2) \\
  UGC4274       &  172.3$\pm$10.9&13.94$\pm$1.20  & 33.3  &  1.4$\pm$0.1   & 1.0$\pm$0.1    & 0.68  & (5) \\  
  IZw18         &  5.0$\pm$0.4   & 0.40$\pm$0.04  & 1.47$\pm$0.09 &  0.04$\pm$0.003& 0.03$\pm$0.003 & 0.03$\pm$0.002 & (6) \\
  VIIZw403      &  [8.4, 11.2]\tablenotemark{a}   & 1.70$\pm$0.14 & 4.90$\pm$0.11  &  [0.07, 0.09]\tablenotemark{a}  & 0.12$\pm$0.01  & 0.10$\pm$0.002 & (7) \\
  Mrk1450       &  47.6$\pm$9.2  & 3.98$\pm$0.74  & 9.55$\pm$0.11 &  0.40$\pm$0.08 & 0.28$\pm$0.06  & 0.19$\pm$0.002 & (8) \\
  UM461         &  13.2$\pm$0.5  & 1.20$\pm$0.04  & 6.10$\pm$0.40 &  0.11$\pm$0.01 & 0.09$\pm$0.003 & 0.12$\pm$0.01 & (9) \\
  Tol1214-277   &  [1.2, 1.6]\tablenotemark{a}    & [0.06,0.12]\tablenotemark{b} & 3.45$\pm$0.10 & [0.010, 0.013]\tablenotemark{a} & [0.004, 0.008]\tablenotemark{b} & 0.07$\pm$0.002 & (10) \\
  Tol65         &  [5.1, 6.7]\tablenotemark{a}    & 0.42$\pm$0.07 & 3.48$\pm$0.10 &  [0.04,0.06]\tablenotemark{a}   & 0.03$\pm$0.002 & 0.07$\pm$0.02 & (10) \\
  Mrk1499       &  38.1$\pm$3.6  & 2.57$\pm$0.30  & 13.2 &  0.32$\pm$0.03 & 0.18$\pm$0.02  & 0.27 & (11) \\
  \enddata
  \tablecomments{We adopt the following values for the solar abundances: (Ne/H)$_\odot$=1.2$\times$10$^{-4}$, (S/H)$_\odot$=1.4$\times$10$^{-5}$ and 
    (O/H)$_\odot$=4.6$\times$10$^{-4}$ (See section 4.1).}
  \tablenotetext{a}{We provide a range for the neon abundance by taking as the lower value the ionic abundance of NeIII, 
    while to obtain the upper value we add the upper limit for NeII.}
  \tablenotetext{b}{We provide a range for the sulfur by taking as the lower value the ionic abundance of SIV,
    while to obtain the upper value we add the upper limit of SIII.}
  \tablerefs{(1) \citet{Bergvall02}, (2) \citet{Guseva00},
    (3) \citet{Izotov06}, (4) \citet{Kobulnicky97}, (5) \citet{Ho97}, 
    (6) \citet{Izotov99a}, (7) \citet{Izotov97}, (8) \citet{Izotov94}, 
    (9) \citet{Izotov98}, (10) \citet{Izotov01}, (11) \citet{Kong02}
    }

\end{deluxetable*}

\subsection{Individual Objects}

Some details on the individual objects and how we measured the
necessary parameters are provided in the following paragraphs.

{\bf Haro11:} The 12.37\,$\mu$m Hu$\alpha$ line is clearly detected in
the SH spectrum.  The source has multiple nuclei, designated as A, B
and C \citep{Bergvall02}.  The optical spectra from which the oxygen
abundance was derived corresponds to the central region. Because all
the lines we are using are from the SH spectrum, we are relatively
free from extinction problems and there is no need for aperture
correction. The contribution of SII may add $\sim$10\% to the total
sulfur abundance.

{\bf NGC1140:} As with Haro11, the 12.37\,$\mu$m Hu$\alpha$ line is
also clearly detected in the SH spectrum of NGC1140.  No measurement
was found in the literature for the electron temperature or density
for this source, thus we assume 10000\,K and 100\,cm$^{-3}$
respectively, which are typical for BCDs. However, because its oxygen
abundance is more than half solar, the electron temperature could also
be lower. If we use a T$_e$ of 5000\,K and re-derive the abundances,
we find that the neon abundance would increase from 5.8$\times$
10$^{-5}$ to 8.5$\times$10$^{-5}$ and the sulfur abundance from
3.9$\times$10$^{-6}$ to 5.5$\times$10$^{-6}$. The contribution of SII
may add $\sim$10\% to the total sulfur abundance.

{\bf SBS0335-052E:} We do not detect the 12.37\,$\mu$m Hu$\alpha$ line
in the SH spectrum.  We use the thermal component of the 5\,GHz radio
continuum from \citet{Hunt04} to convert to the H$\beta$ flux, and
thus no extinction correction is needed. [NeII] is not detected and we
provide the upper limit in Table \ref{tab3}. The ionic abundance of
SII is $\sim$25\% of the SIII determined from the optical study of
\citet{Izotov06} and may add 12\% to the total sulfur abundance.  We
also find higher neon and sulfur abundances compared with the oxygen
and possible implications are discussed in \citet{Houck04b}.

{\bf NGC1569:} The 12.37\,$\mu$m Hu$\alpha$ line is clearly detected
in the SH spectrum, thus providing a direct estimate of the H$\beta$
flux inside the SH slit. This galaxy is extended in the mid-IR
\citep[see Fig. 4 in][]{Wu06} and optical spectra have been taken for
several of the bright knots by \citet{Kobulnicky97} in order to study
the chemical gradient and inhomogeneities of the source. These authors
found very little variation in the metallicity. Our infrared-derived
abundances are in rough agreement with the optical results. The ionic
abundance of SII is $\sim$13\% that of the SIII estimated from the
optical study \citep{Kobulnicky97} and may add 9\% to the total sulfur
abundance.

{\bf IIZw40:} We use the 12.37\,$\mu$m Hu$\alpha$ line to convert to
H$\beta$ flux. The derived neon and sulfur abundances agree with each
other but appear to be lower than oxygen with respect to the solar
values. If we estimate the H$\beta$ flux from the H$\alpha$ image of
\citet{Gildepaz03}, we find a value $\sim$22\% lower than the first,
which would increase the neon and sulfur abundances by 22\%
accordingly. This suggests, that when using the H$\beta$ flux derived
from H$\alpha$ image, the systematic error in our measurement is
probably no better than 25\%.  The ionic abundance of SII is
$\sim$13\% that of SIII from the optical study of \citet{Guseva00} and
may add 7\% to the total sulfur abundance.

{\bf UGC4274:} We use an H$\beta$ flux derived from the H$\alpha$
image of \citet{Gildepaz03}. Similarly to NGC1569, the galaxy is
extended and the peak of the infrared centroid is displaced from the
optical peak position. The derived neon and sulfur abundances are both
super solar. The oxygen abundance is not directly available from the
literature, thus we use the [NII]/H$\alpha$ method \citep{Ho97,
  Denicolo02} to derive O/H, which bears a large uncertainty. The
contribution of SII may add $\sim$10\% to the total sulfur abundance.

{\bf IZw18:} The H$\beta$ flux is derived from the thermal component
of the radio continuum inside the SH slit.  The ionic abundance of SII
is $\sim$21\% that of SIII from the optical study of \citet{Izotov99a}
and may add 15\% to the total sulfur abundance. A more detailed
discussion on this object can be found in \citet{Wu07}.

{\bf VIIZw403:} The H$\beta$ flux is derived by using the H$\alpha$
image from \citet{Gildepaz03}.  The optical and infrared centroids do
not overlap. The SH spectrum for this source is noisy and
[NeII]\,$\lambda$12.81\,$\mu$m is not detected, thus we have only a lower limit
on the neon abundance.  The upper limit of the [NeII] line indicates
that it could add less than 25\% to the total elemental abundance of
neon which is presented in Table \ref{tab4}. The ionic abundance of
SII is $\sim$24\% that of SIII from the optical work of
\citet{Izotov97} and may add 20\% to the total sulfur abundance.

{\bf Mrk1450:} As with VIIZw403, we use the hydrogen flux as given by
the H$\alpha$ flux included in the SH slit.  Neon and sulfur
abundances appear to be higher than the oxygen abundance with respect
to the solar values. The ionic abundance of SII is $\sim$18\% that of
SIII from the optical study of \citet{Izotov94} and may add 12\% to
the total sulfur abundance.

{\bf UM461:} The H$\beta$ flux is estimated from the integrated
spectra of \citet{Moustakas06}, and the source is small enough to be
fully included by the SH slit. The derived neon and sulfur abundances
are in good agreement with oxygen metallicity from the optical. The
ionic abundance of SII is $\sim$17\% that of SIII from the optical
\citep{Izotov98} and may add 8\% to the total sulfur abundance.

{\bf Tol1214-277:} The 12.37\,$\mu$m Hu$\alpha$ line is marginally
detected in the SH spectrum, and we use this line to derive the
H$\beta$ flux. The spectrum is very noisy and no
[NeII]\,$\lambda$12.81\,$\mu$m or [SIII]\,$\lambda$18.71\,$\mu$m lines
are visible. We use the ionic abundances of [NeIII] and [SIV] as the
lower limits of the neon and sulfur abundances.  The upper limit of
[NeII] amounts to 33\% of the ionic abundance of [NeIII] while the
upper limit of [SIII] is nearly as much as the abundance of [SIV].  We
give the range of their elemental abundances in Table \ref{tab4}. The
ionic abundance of SII is $\sim$18\% that of SIII from the optical
study of \citet{Izotov01} and may add $\sim$9\% to the total sulfur
abundance.

{\bf Tol65:} We use the H$\alpha$ image \citep{Gildepaz03} to
calculate the H$\beta$ flux. [NeII]\,$\lambda$12.81\,$\mu$m is not
detected in the SH spectrum, thus our derived neon abundance is a
lower limit. The upper limit of NeII is $\sim$32\% of the ionic
abundance of NeIII, and the range of neon abundance is provided in
Table \ref{tab4}.  The metallicity of sulfur relative to solar is
lower than oxygen. The ionic abundance of SII is $\sim$26\% that of
SIII from the optical \citep{Izotov01} and may add 19\% to the total
sulfur abundance.

{\bf Mrk1499:} The 12.37\,$\mu$m Hu$\alpha$ line is detected in the SH
spectrum. There was a bad pixel at the peak of this line and we
interpolated its value using adjacent pixel values. The infrared
centroid appears to be slightly shifted from the optical one. The
H$\beta$ flux inside the SH slit derived from the H$\alpha$ image is
very similar to that derived from the Hu$\alpha$ line (within
$\sim$10\%). The contribution of SII may add $\sim$10\% to the total
sulfur abundance.

\section{Discussion}

In this section, we compare the abundances derived from the IR spectra
with the results from the optical studies.

\subsection{Solar abundances}

The solar photospheric abundances have changed drastically over the
past few years \citep[see discussion in][]{Pottasch06}. In some cases,
the solar abundance estimates for the same element from different
authors could differ by a factor of two. In this paper, when we quote
the elemental abundance in solar units (e.g. Table \ref{tab4}), we
adopt the following values: (Ne/H)$_\odot$=1.2$\times$10$^{-4}$ from
\citet{Feldman03}, (S/H)$_\odot$=1.4$\times$10$^{-5}$ from
\citet{Asplund05}, and (O/H)$_\odot$=4.9$\times$10$^{-4}$ from
\citet{Prieto01}. We should note though that \citet{Asplund05} have
also reported that (Ne/H)$_\odot$=6.9$\times$10$^{-5}$ and
\citet{Grevesse98} have given (S/H)$_\odot$=2.1$\times$10$^{-5}$,
while \citet{Anders89} have reported that
(O/H)$_\odot$=8.5$\times$10$^{-4}$. These values represent the
extremes for the solar neon, sulfur and oxygen abundance
determinations.  \citet{Pottasch06} have shown that the higher neon
solar value is favored in their sample of planetary nebulae (PNe). In
the present paper we indicate the range of solar abundances in our
plots and we also discuss the impact of different solar values on our
results.

\subsection{Neon and sulfur abundances}

As discussed in \citet{Thuan95}, it is assumed --- and has been shown
in the optical --- that the $\alpha$-element-to-oxygen abundance
ratios do not vary with oxygen abundance. This is due to the fact that
those elements are produced by the same massive stars
(M$>$10M$_{\odot}$) responsible for oxygen production. We test this
assumption in the infrared by studying the abundances of neon and
sulfur as derived from our IRS data.

\begin{figure}
  \epsscale{1.2}
  \plotone{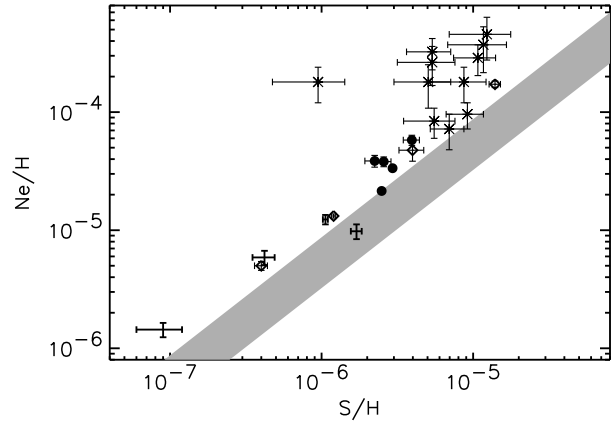}
  \caption{A plot of the Ne/H vs S/H abundance. Our BCDs are indicated by the
    filled circles if Hu\,$\alpha$ is detected and by diamonds if it is not detected. The sources 
    that are shown by only the error bars are those that have non-detection of [NeII] 
    or [SIII] (See note of Table \ref {tab4}). The starburst galaxies from \citet{Verma03} 
    are marked with the stars. We use the grey band to indicate the locus on the plot where the 
    ratio of Ne/S would be consistent with the different solar values for neon and sulfur abundances.}
  \label{fig:fig2}
\end{figure}

In Fig. 2, we plot the abundances of neon and sulfur as derived from
our high-resolution data. \citet{Verma03} have found a positive
correlation between the neon and argon abundances for their sample of
starburst galaxies while their data show no correlation between the
sulfur and neon and/or argon abundances (indicated as stars on Fig.
2). However, our sources show that the neon and sulfur abundances
scale with each other. In the same figure, we also plot the
proportionality line of the ratio for (Ne/S)$_\odot$. The maximum and
minimum values of the solar neon and sulfur abundances are indicated
by the width of the grey band, and we find that most of our BCDs have
ratios above those values.

\begin{figure}
  \epsscale{1.2}
  \plotone{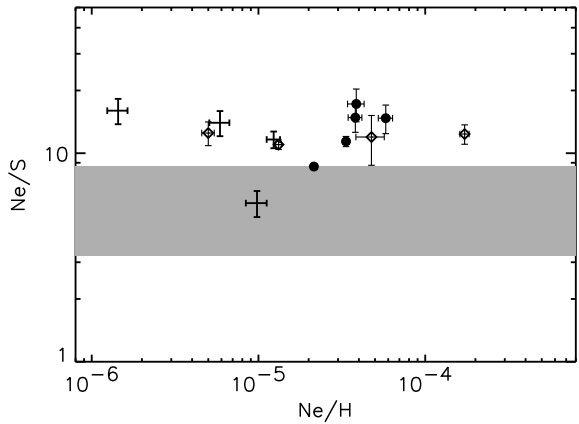}
  \caption{The abundance ratio of Ne/S as a function of Ne/H. The
    symbols are the same as in Fig. 1. The grey area indicates the
    range for the values of the Ne/S ratio encountered in the solar
    neighborhood.}
  \label{fig:fig3}
\end{figure}

In Fig. 3, we examine the ratio of Ne/S as a function of Ne/H derived
from the IRS data. We see no correlation between the ratio of Ne/S
with respect to Ne/H, in agreement with the results of
\citet{Thuan95}.  The average Ne/S ratio we found for these BCDs is
$\sim$11.4. This is similar with the Ne/S ratio of 14.3 in the Orion
Nebula \citep{Simpson04}.  The Ne/S ratios found in the HII regions of
M83 are higher ($\sim$24$-$42) and decrease with increasing
deprojected galactocentric radii \citep{Rubin07}.  These authors have
also found that the Ne/S ratios in M33 are $\sim$12$-$21
\citep{Rubin06}. We notice that all the BCDs, as well as the HII
regions in M33, M83 and the Orion Nebula, have larger Ne/S ratios than
those found in the solar neighborhood, while their Ne/S ratios from
optical measurements are more consistent with the solar ratios.  The
higher Ne/S ratios from infrared studies have already been found in
other type of objects, including PNe, starburst galaxies, and other
HII regions \citep{Marigo03,Verma03, Henry04,Pottasch06}. Some of
these works point out that the solar sulfur abundance could have been
underestimated and they also appear to favor the higher values for the
solar neon abundance.  It could also be due to a differential
depletion of sulfur onto dust grains compared to neon as suggested by
several studies \citep{Simpson90,Verma03,Pottasch06, Bernard-Salas07}.

\begin{deluxetable*}{lcccccccc}
  \tabletypesize{\scriptsize}
  \setlength{\tabcolsep}{0.02in}
  \tablecaption{PAH measurements of the sample\label{tab5}}
  \tablewidth{0pc}
  \tablehead{
    \colhead{Object Name} & \multicolumn{4}{c}{PAH EW ($\mu$m)} & \multicolumn{4}{c}{Integrated Flux ($\times10^{-14}$ergs~cm$^{-2}$~s$^{-1}$})\\
    \colhead{} & \colhead{6.2\,$\mu$m} & \colhead{7.7\,$\mu$m} & \colhead{8.6\,$\mu$m} & 
    \colhead{11.2\,$\mu$m} &  \colhead{6.2\,$\mu$m} & \colhead{7.7\,$\mu$m} & \colhead{8.6\,$\mu$m} & 
    \colhead{11.2\,$\mu$m} \\
  }
  \startdata
 Haro11      &   0.120$\pm0.003$  &  0.227$\pm0.005$  &  0.042$\pm0.003$  &   0.099$\pm0.002$  &  78.3$\pm1.8$     &    180$\pm4$       & 25.4$\pm1.6$      &   67.2$\pm1.3$ \\ 
 NGC1140     &   0.479$\pm0.033$  &  0.568$\pm0.024$  &  0.091$\pm0.006$  &   0.525$\pm0.032$  &  42.1$\pm1.4$     &   78.3$\pm1.8$     & 11.1$\pm0.7$      &   50.2$\pm2.5$ \\
 SBS0335-052E&$<$0.035            & \nodata           & \nodata           &$<$0.015            &  $<$1.7           &  \nodata           & \nodata           &   $<$1.0 \\
 NGC1569     &   0.232$\pm0.013$  &  0.380$\pm0.019$  &  0.007$\pm0.005$  &   0.129$\pm0.008$  &  50.4$\pm2.4$     &    118$\pm6$       &  2.7$\pm1.6$      &   68.1$\pm4.3$ \\
 IIZw40      &   0.044$\pm0.007$  &  0.044$\pm0.006$  &  0.006$\pm0.005$  &   0.033$\pm0.006$  &  13.2$\pm2.1$     &   19.0$\pm2.8$     &  2.8$\pm1.2$      &   21.1$\pm3.7$ \\
 UGC4274     &   0.423$\pm0.032$  &  0.497$\pm0.049$  &  0.105$\pm0.012$  &   0.495$\pm0.021$  &  23.2$\pm1.2$     &   43.3$\pm2.3$     &  7.7$\pm0.9$      &   29.1$\pm0.9$ \\ 
 IZw18       &$<$0.233            & \nodata           & \nodata           &$<$0.116            &  $<$0.4           &  \nodata           & \nodata           &   $<$0.1 \\  
 Mrk1450     &   0.207$\pm0.045$  &  0.337$\pm0.143$  &  0.084$\pm0.020$  &   0.112$\pm0.017$  &   1.8$\pm0.4$     &    2.8$\pm0.9$     &  0.7$\pm0.2$      &    1.6$\pm0.2$ \\ 
 UM461       &$<$0.663            & \nodata           & \nodata           & $<$0.134           &  $<$1.5           &  \nodata           & \nodata           &   $<$1.6 \\
 Mrk1499     &   0.408$\pm0.040$  &  0.526$\pm0.016$  &  0.156$\pm0.011$  &   0.721$\pm0.035$  &    5.5$\pm0.3$    &   10.9$\pm0.3$     &  2.3$\pm0.2$      &    7.3$\pm0.2$ \\

 \enddata \tablecomments{Contrary to \citet{Wu06}, the wavelengths
   which we have chosen to determine the underlying continuum are
   fixed in the above measurements, which explains the minor
   discrepancies between these two studies. The symbol `` \nodata''
   indicates that no PAH EW measurement was possible and it is mostly
   due to the low SNR of the corresponding spectrum. In this case, the
   determination of the continuum is highly uncertain and could
   significantly affect the value of the PAH EW. For SBS0335-052E, the
   SNR is high enough, but no PAH features can be identified in its
   mid-IR spectrum. We do not derive upper limits for the PAH EWs at
   7.7\,$\mu$m and 8.6\,$\mu$m, because no reliable templates for
   those features are available.}
  
\end{deluxetable*}

\subsection{Comparison with optically derived abundances}

As has been discussed earlier in this paper, one of the major
advantages of using IR lines for estimating abundances is that one can
probe emission from dust enshrouded regions without the uncertainties
introduced from extinction corrections.  In this subsection, we
compare the newly derived neon and sulfur abundances of our BCD sample
using the infrared data, with the abundances of the same elements as
well as those for oxygen estimated from optical studies available in
the literature.

In order to determine elemental abundances, the optical studies
usually adopt a two-zone photoionized HII region model: a
high-ionization zone with temperature T$_e$(OIII) and a low-ionization
zone with temperature T$_e$(OII). Because the infrared lines are much
less sensitive to variations in T$_{e}$ (which affect the calculation
on the population of the various atomic levels), in the subsequent
calculations, we assume a constant electron temperature for our
analysis.

\begin{figure}
  \epsscale{2.2}
  \plottwo{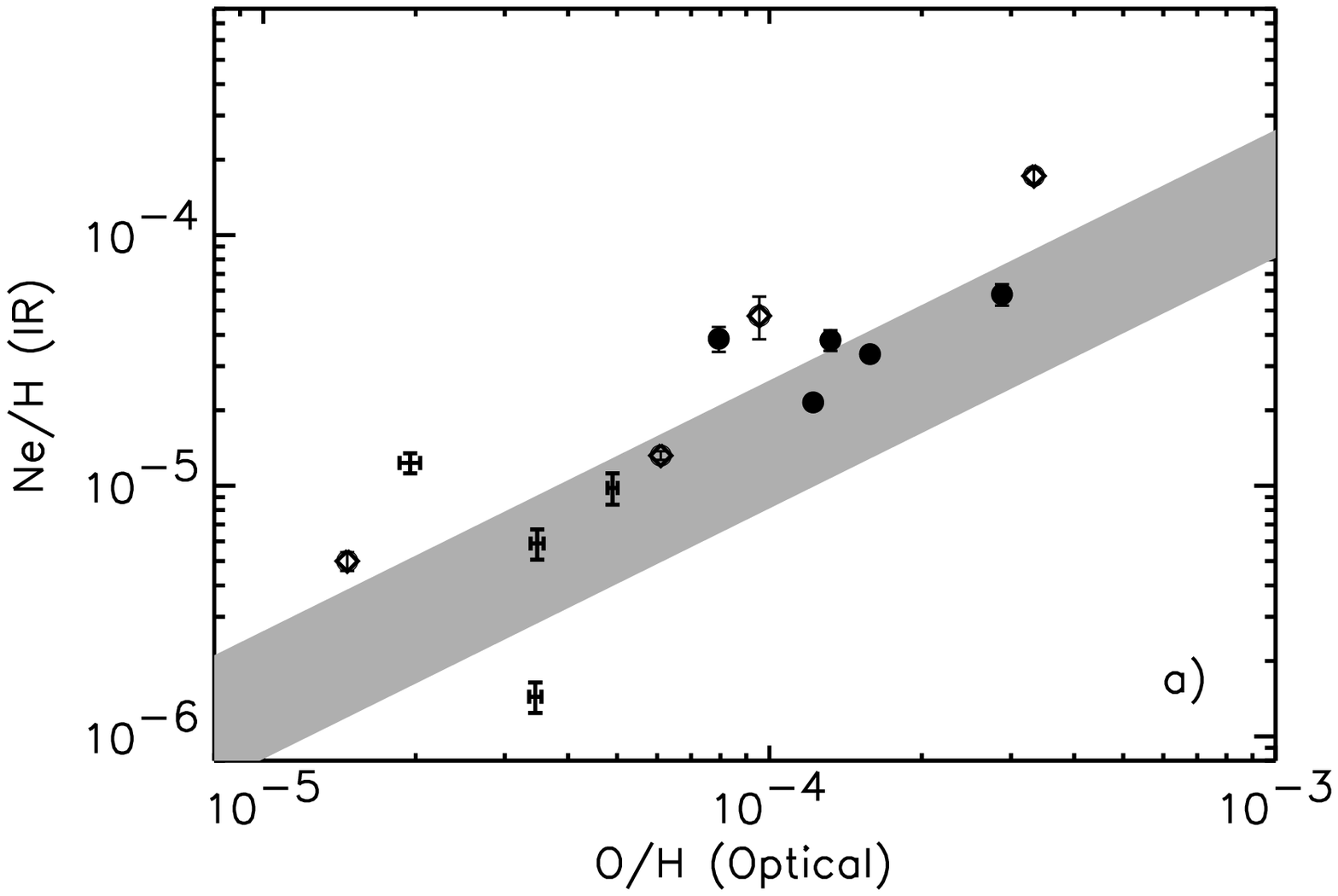}{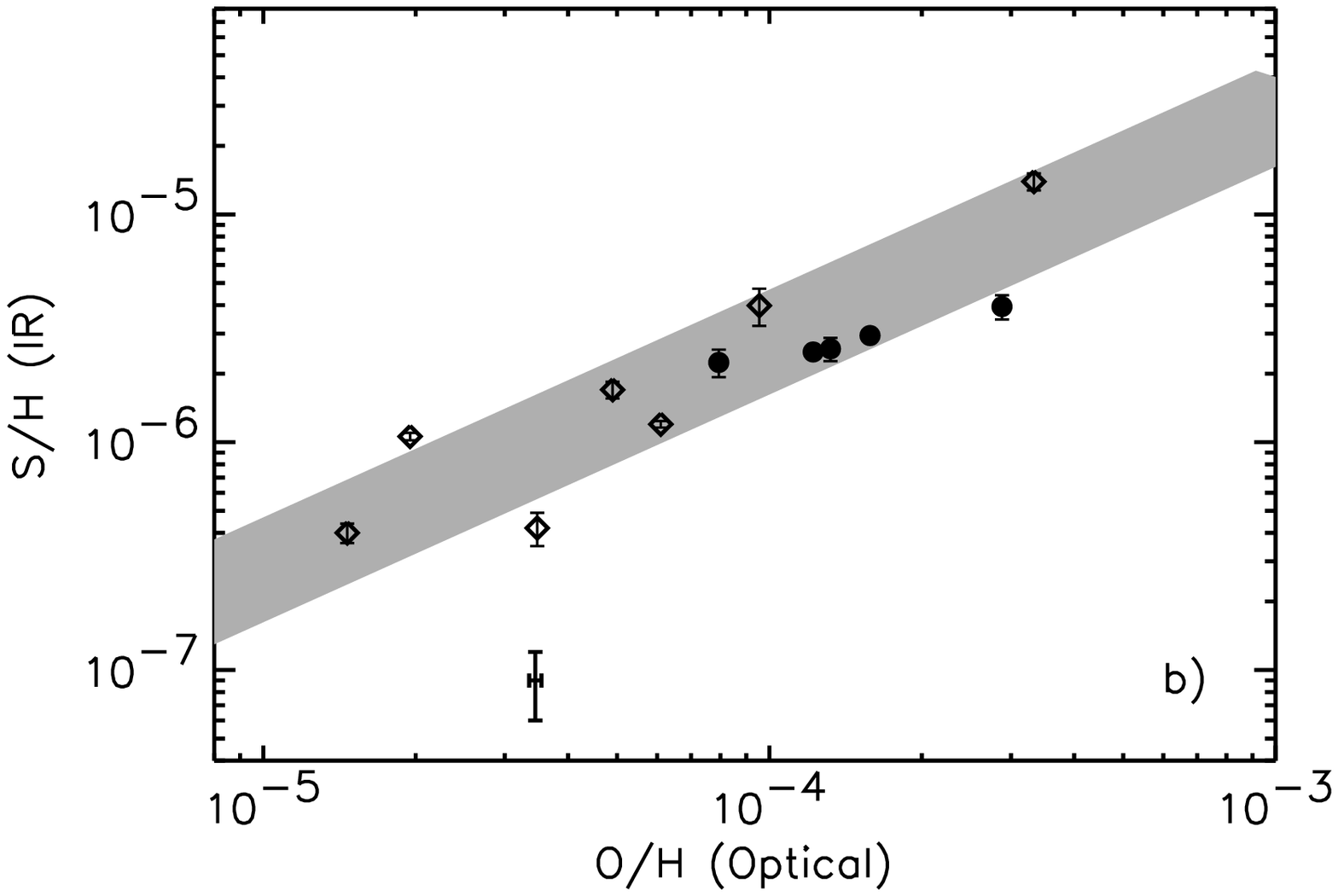}
  \caption{a)The Ne/H vs O/H abundances. The symbols are the same as in Fig. 1. The grey band 
    represents the locus on the plot where sources of different metallicities would have a Ne/O
    intrinsic ratio similar to the values found in the solar neighborhood. 
    b) Same as a), but for S/H vs O/H abundances.}
  \label{fig:fig4}
\end{figure}

In Fig. 4 we compare the infrared derived neon and sulfur abundances
of the BCD sample with their optical metallicities, using the oxygen
abundances from the literature (see also Table \ref{tab4}).  The
infrared neon abundances with respect to solar are slightly higher
than the oxygen (see Fig. 4a), but overall, there is a good agreement
between the infrared measurements and the optical results. This
indicates that there is little dust enshrouded gas, or that these
regions have similar metallicities, or that they do not contribute much
to the total integrated emission of the galaxies. In SBS0335-052E, as
well as Mrk1450, UGC4274 and Haro11, the infrared derived metallicity
of neon is more than twice that of oxygen compared to the solar
values.  This could be due to the different regions we are probing
using the infrared, which could be obscured in the optical.
SBS0335-052E is known to have an embedded super star cluster that is
invisible in the optical \citep[see][]{Houck04b}. Both Haro11 and
UGC4274 are somewhat extended, and detailed spectroscopy over the
whole galaxy would be needed to directly compare the infrared results
with the optical ones. For Mrk1450, which is a point source for
Spitzer/IRS, the abundance we derived using the infrared data is
higher than the optical value and it is likely due to the fact that
the optical estimates are more susceptible to the uncertainties in the
electron temperature. If we lower T$_e$ from 12\,500\,K to 10\,000\,K,
the oxygen abundance calculated from the optical data would double,
while the neon abundance measured from the infrared would only
increase by 5\%. In Fig. 4b we show that the infrared derived sulfur
abundances agree well with the oxygen abundance relative to solar.

For nine out of the thirteen objects we studied, optically derived
neon and sulfur abundances are also available in the literature. The
optical data have matched apertures, in most cases an accurate T$_e$
measurement, a direct measure of the H$\beta$ flux, and reddening
derived from the same data.  Our infrared results are less sensitive
to the electron temperature, but affected by the uncertainty in the
H$\beta$ flux if the Hu\,$\alpha$ line is not detected and the
uncertainty could be as large as 25\%.  In Fig. 5a, we plot the
optically derived Ne/H against the infrared Ne/H. The solid line is
the 1:1 proportion line for the infrared and optical derived
abundances and we find that most of the sources are located above this
line. A possible explanation for the higher infrared derived neon
abundances could be the presence of dust enshrouded regions which
might have higher heavy element abundances. This could explain the
higher infrared metallicity found in SBS0335-052E, but if this were
true for the whole sample, it should also apply to the sulfur
abundances.  However, when we plot the S/H (Optical) against the S/H
(IR) in Fig. 5b, we find that the sources are located on both sides of
the 1:1 proportion line. Alternatively, this could be due to the
difference in the determination of element abundance using the
infrared and optical methods. Because only NeIII is detected in the
optical regime, the total elemental abundance of neon from the optical
study is heavily dependent on the adopted ICF. In the infrared the
[NeII] line is detected in most of our sources and we do not use any
ICF for our study. Thus the higher neon abundances derived from the
infrared data compared to the optical results could also be due to the
large uncertainty in the ICF used in the optical studies. Another
possibility is that the temperature of the NeIII ion is lower than
that of the OIII ion as found in some PNe \citep{Bernard-Salas02}.
Because the optical studies are based on T$_e$(OIII) when calculating
the ionic abundance of NeIII, that could also result in an
underestimate of its abundance.

\begin{figure}
  \epsscale{2.2}
  \plottwo{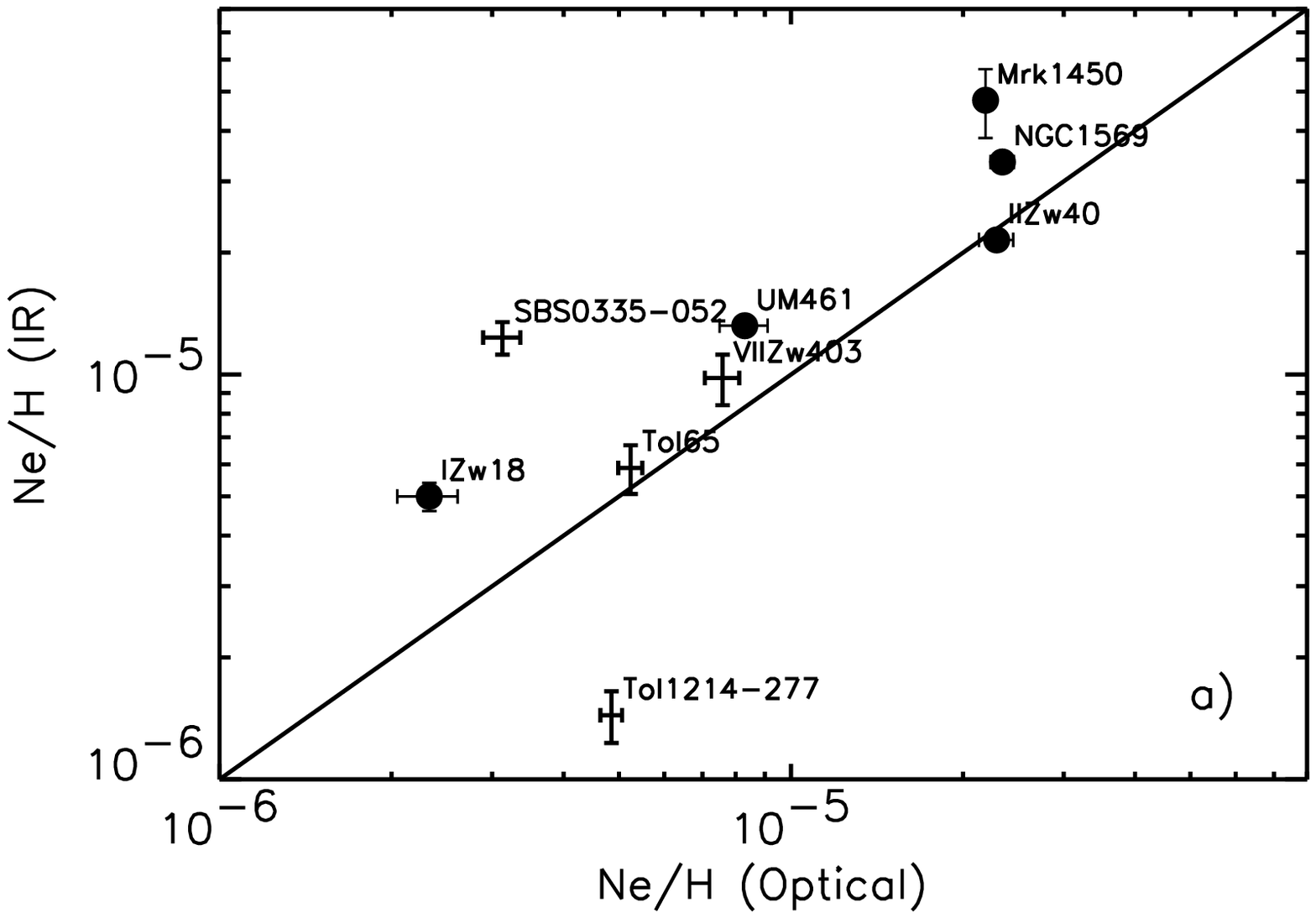}{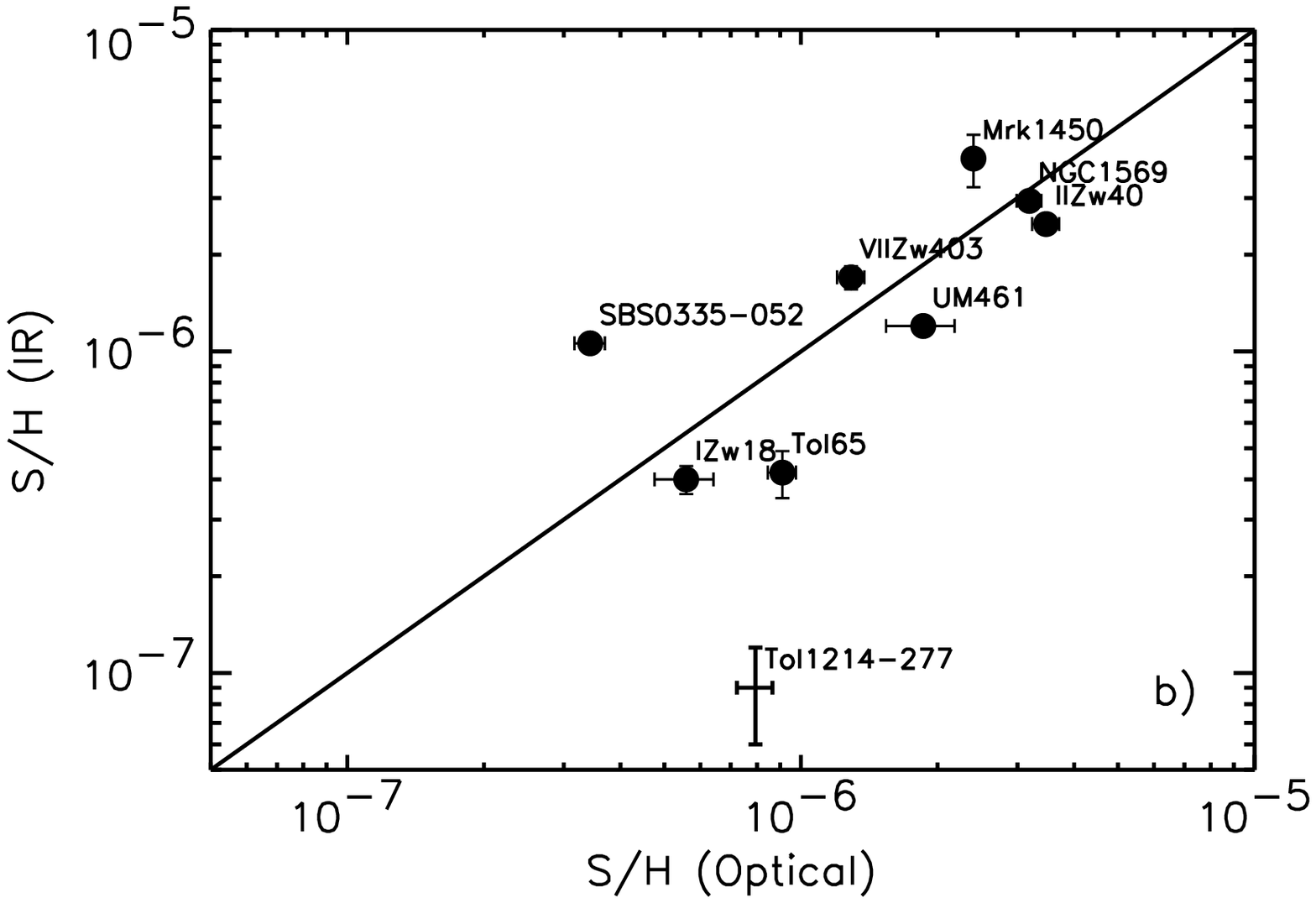}
  \caption{a)The infrared derived Ne/H as a function of the optical
    Ne/H abundances. The sources shown with only the error bars are
    those where [NeII] or [SIIII] is not detected. The solid line 
    represents the 1:1 line for the optical and IR derived abundances.  
    b) Same as a), but for S/H (Optical) vs S/H (IR) abundances.}
  \label{fig:fig5}
\end{figure}

Finally, we explore the variation in the ratio of Ne/S as derived from
the infrared and the optical data, and plot them as a function of O/H
in Fig. 6. Interestingly, we find that with the exception of VIIZw403,
in all BCDs, the infrared derived Ne/S ratios are higher than the
corresponding optical values. \citet{Izotov99b} indicate an average
Ne/S ratio of 6.9 for the 54 super giant HII regions they study in 50
BCDs using optical spectroscopy. The average of the optically derived
Ne/S ratio for the nine sources in our sample is 6.5$\pm$1.8 while the
corresponding ratio from the infrared data is 11.4$\pm$2.9,
statistically higher than the optical results. As we mentioned
earlier, no SII abundances could be determined with the Spitzer/IRS
data.  However, in high-excitation objects such as BCDs, SII does not
contribute much ($\sim$10\%) to the total elemental abundance, thus it
could not account for the factor of two difference in the ratio of
Ne/S.  Furthermore, such a discrepancy is not a result of metallicity
because we see no correlation in the dispersion between the infrared and
optically derived ratios of Ne/S with respect to O/H. If we divide the
sources into a high-excitation group with sources that have a
detectable [OIV]$\lambda$25.89\,$\mu$m line and a low-excitation group
with sources without any detection of [OIV], we still see no clear
trend. Thus the ionization field cannot explain the difference in the
Ne/S ratios either. Moreover, no correlation is found between the Ne/S
ratio and the extinction to the source. Even though the largest
uncertainty in the infrared derived abundances comes from the
uncertainty in the H$\beta$ flux in cases where no Hu\,$\alpha$ is
detected, when calculating the ratio of Ne/S, the H$\beta$ flux
cancels out, thus our infrared determined values for Ne/S should be
fairly reliable. One possibility for the observed discrepancy could be
that the difference in the infrared and optical results is again due
to the ICF adopted in the optical studies.

\begin{figure}
  \epsscale{1.2}
  \plotone{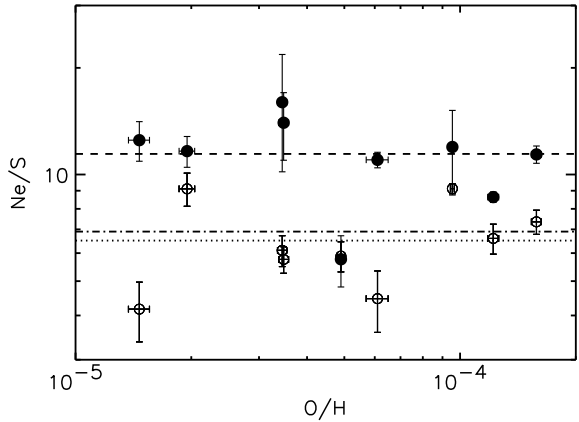}
  \caption{The abundance ratio of Ne/S as a function of O/H. The infrared derived Ne/S ratios are 
    indicated by the filled circles while the optical derived Ne/S ratios are shown by the open 
    circles. The dashed line is the average Ne/S ratio from the infrared derived elemental abundances 
    while the dotted line is average Ne/S ratio from the optical derived elemental abundances. 
    The dash-dotted line represents the average Ne/S ratio from \citet{Izotov99b}. }
  \label{fig:fig6}
\end{figure}

\section{Metallicities and PAHs}

The observed suppression of PAH emission in low metallicity
environment is still an open topic \citep{Madden00,Madden06,
  Engelbracht05, Wu06, OHalloran06}.  One explanation on the absence
of PAHs is that there is not enough carbon to form into PAH molecules
in such metal-poor environments while an alternative possibility could
be that the strong ionization field in low metallicity environments
may destroy the fragile PAH rings.

Previous studies relating the PAHs and metallicity were all based on
the oxygen abundance measured from the optical, which could be
associated with regions that are different from the ones where PAH
emission originates. Because we have obtained the infrared measured neon
and sulfur abundances, it is particularly interesting to compare the
PAH strength with our metallicity estimates. Several BCDs in our
sample do show PAH emission features and an analysis of the PAH
properties of our sample based on IRS low resolution spectra was
presented in \citet{Wu06}. We have obtained deeper spectra for some of
the sources, so for consistency, we have re-measured all features in
our sample and the results are given in Table \ref{tab5}.  None of the
sources with metallicity lower than $\sim$0.12\,Z$_\odot$ show any
detectable PAHs. For the sources where PAHs were detected, we
calculate the equivalent widths (EWs) of the 6.2,7.7,8.6 and
11.2\,$\mu$m features and find that the correlation between the PAH
EWs and the metallicity is weak. This is not unexpected because as
discussed in \citet{Wu06}, the strength of PAHs in low-metallicity
environments is a combination of creation and destruction effects. As
a result the dependency of PAH strength on the metallicity alone could
have a substantial scatter.

\section{Conclusions}

We have studied the neon and sulfur abundances of thirteen BCDs using
Spitzer/IRS high-resolution spectroscopy.  Our analysis was based
on the fine-structure lines and the hydrogen recombination line
detected in the SH spectra of the {\em IRS}, combined with the radio
continuum, H$\alpha$ images and integrated optical spectral data in
some cases. We find a positive correlation between the neon and sulfur
abundances, though sulfur appears to be more under-abundant than neon
(with respect to solar). The ratio of Ne/S for our sources is on
average 11.4$\pm$2.9, which is consistent with what has been found in
other HII regions using infrared data. However, this average ratio
appears to be higher than the corresponding optical value of
6.5$\pm$1.8 (in BCDs), which could be due to the adopted ICFs in the
optical studies. When comparing the newly derived neon and sulfur
abundances with the oxygen abundances measured from the optical lines,
we find a good overall agreement.  This indicates that there are few
completely dust enshrouded HII regions in our BCDs, or if such HII
regions are present, they have similar metallicities to the ones
probed in the optical. Finally, the infrared derived neon and sulfur
abundances also correlate, with some scatter, with the corresponding
elemental abundances derived from the optical data.

\acknowledgments We thank Robert Kennicutt whose detailed comments
help to improve this manuscript. We also thank Shannon Gutenkunst and
Henrik Spoon for helpful discussions as well as an anonymous referee
whose insightful suggestions helped to improve this manuscript. This
work is based in part on observations made with the Spitzer Space
Telescope, which is operated by the Jet Propulsion Laboratory,
California Institute of Technology, under NASA contract 1407. Support
for this work was provided by NASA through Contract Number 1257184
issued by JPL/Caltech.

\acknowledgments

\end{document}